\documentclass[a4paper,12pt]{article}
\newcommand{\lesssim}{\:\mbox{\raisebox{-3pt}{$\stackrel%
{\displaystyle <}{\sim}$}}\:}
\begin{document}
\title{
\begin{flushright}
{\small\begin{tabular}{r} UWThPh-2001-23\\ July 2001
\end{tabular}}\\
\end{flushright}
\vspace{1cm}
Bell inequality and $CP$ violation in the neutral kaon system}

\vspace{2cm}

\author{R.A. Bertlmann\footnote{E-mail: bertlman@ap.univie.ac.at}\hspace{4pt},
\addtocounter{footnote}{1}
W. Grimus\footnote{E-mail: grimus@doppler.thp.univie.ac.at} 
\addtocounter{footnote}{3}
and B.C. Hiesmayr\footnote{E-mail: hies@thp.univie.ac.at}\\
{\em Institute for Theoretical Physics, University of Vienna}\\ {\em
Boltzmanngasse 5, 
A-1090 Vienna, Austria}}

\date{}

\maketitle

\begin{abstract}
For the entangled neutral kaon system we formulate a Bell inequa\-lity
sensitive to $CP$ 
violation in mixing. Via this Bell inequality we ob\-tain a
bound on the leptonic $CP$ asymmetry which is violated by experimental
data. Furthermore, we connect the Bell inequality with a decoherence
approach and find a lower bound on the 
decoherence parameter which practically corresponds to Furry's hypothesis.

\vspace{0.5cm}

\noindent PACS numbers: 03.65.Ta, 03.65.Ud, 13.25.Es, 11.30.Er\\
Key words: Bell inequality, nonlocality,
entangled states, neutral kaons, $CP$ violation
\end{abstract}

\vspace{1cm}

Recently there has been great interest in investigating entangled
massive particle systems, e.g., neutral kaons
\cite{Bramon,ghirardi91,trixi,BH1,BGH2}. In analogy to spin 
$1/2$ particles or to polarized photons \cite{Zeilinger}, neutral
kaons also can be 
described by a ``quasi spin'', a view which is especially useful in
this connection (see, e.g., Ref. \cite{BH1}). They are ideal systems
to test the EPR-Bell correlations for massive 
systems. A general test of quantum mechanics (QM) versus local realistic theories (LRT)
is performed via Bell inequalities \cite{bell2}. In the kaon case we have the freedom of
choosing different detection times and different quasi spins. They
play the role of the different angle choices in the spin $1/2$ or
photon case. Experimentally such 
systems are produced at the $\Phi$ resonance, for instance, in the
$e^+ e^-$ collider
DA$\Phi$NE at Frascati or in $p \bar p$ collisions in the CPLEAR
experiment at CERN.

An interesting feature of the kaon systems is that the kaons reveal $CP$ violation and,
amazingly, it turns out that Bell inequalities for such systems imply bounds on the
physical $CP$ violation parameters \cite{Uchiyama,Benatti}, which can be checked
experimentally, indeed, not necessarily in experiments with entangled kaons.

It was Uchiyama \cite{Uchiyama} who first found that a Bell inequality with different
quasi spin eigenstates leads to an inequality for the $CP$ violation parameter
$\varepsilon \,$. The derivation relied on a specific phase convention for the
kaon states. Such a specific choice, although customary in kaon
physics, is a certain
drawback for the physical interpretation of Bell inequalities since their formulation
should be as general and loophole free as possible (see, e.g.,
Ref. \cite{Genovese}). 

It is the purpose of the present paper to optimize the Bell inequality
(BI) for such entangled kaons by exploiting the phase freedom in the
definition of the kaon states. In this way we will clarify the
relation between Uchiyama's BI and CP violation in mixing.

Quantum mechanically we are considering entangled states of $K^0 \bar K^0$
pairs, in analogy to the entangled spin up and down pairs, or photon pairs.
They are created through the reaction $e^+ e^- \to \Phi \to K^0 \bar K^0$ in
a $J^{PC}=1^{--}$ quantum state, and thus antisymmetric under $C$
and $P$, and are described at the time $t=0$ by the entangled state
\begin{eqnarray}\label{entangledK0}
| \psi (t=0) \rangle &=&\frac{1}{\sqrt{2}}
\left\{ | K^0 \rangle_l \otimes\! | \bar K^0 \rangle _r -
| \bar K^0 \rangle _l \otimes | K^0 \rangle _r \right\},
\end{eqnarray}
which can be rewritten in the $K_S K_L$ basis as
\begin{eqnarray}\label{entangledKS}
| \psi (t=0) \rangle&=&
 \frac{N_{SL}}{\sqrt{2}}\left\{ | K_S \rangle_l \otimes\! | K_L \rangle _r -
| K_L \rangle _l \otimes | K_S \rangle _r \right\}
\end{eqnarray}
with $N_{SL}=N^2/(2pq)$. 
Then the neutral kaons fly apart and will be detected on
the left ($l$) and right ($r$) side of the source. Of course, during their propagation
the $K^0$ and $\bar K^0$ oscillate and $K_S, K_L$ decays will occur.

\vspace{6mm}

What is Uchiyama's inequality? Imagine the following gedanken experiment. Two neutral
kaons are produced at the $\Phi$ resonance, each one in a definite quasi spin state. The
probability of measuring the short lived state $K_S$ on the left
side and the 
anti-kaon $\bar K^0$ on the right side, at the time $t=0$, is denoted
by $P(K_S,\bar 
K^0)$, and analogously the probabilities $P(K_S,K_1^0)$ and $P(K_1^0,\bar K^0)$. Then
under the usual hypothesis of Bell's locality the following Bell inequality can be
derived \cite{Uchiyama}: 
\begin{equation}\label{UchiyamaBI}
P(K_S,\bar K^0)\; \leq\; P(K_S,K_1^0) + P(K_1^0,\bar K^0) \, .
\end{equation}
Generalizations can be found in Ref. \cite{BH1}.
Although this BI is rather formal because it involves the unphysical $CP$-even
state $| K^0_1 \rangle$, it implies an inequality on the physical
$CP$ violation parameter $\varepsilon \,$, which is experimentally testable.
The procedure to derive this inequality is as follows.

In QM we describe the neutral kaons by three sets of quasi spin eigenstates.
Let us begin with the strangeness eigenstates. They distinguish the $K^0$
from its antiparticle $\bar K^0$ by
\begin{eqnarray}
S|K^0\rangle \; &=& + |K^0\rangle \,, \nonumber\\
S|\bar K^0\rangle \; &=& - |\bar K^0\rangle \, .
\end{eqnarray}
As the $K$ mesons are pseudoscalars, their parity $P$ is negative and
charge conjugation 
$C$ transforms $K^0$ and $\bar K^0$ into each other so that we
conventionally have for the combined transformation $CP$
\begin{eqnarray}\label{CPtransform}
CP|K^0\rangle \; &=& - |\bar K^0\rangle \,, \nonumber\\
CP|\bar K^0\rangle \; &=& - |K^0\rangle \, .
\end{eqnarray}
From this follows that the orthogonal linear combinations
\begin{eqnarray}
|K_1^0\rangle \; &=& \frac{1}{\sqrt{2}}\big\lbrace |K^0\rangle -
|\bar K^0\rangle \big\rbrace \,, \nonumber\\
|K_2^0\rangle \; &=& \frac{1}{\sqrt{2}}\big\lbrace |K^0\rangle +
|\bar K^0\rangle \big\rbrace
\end{eqnarray}
are eigenstates of $CP$
\begin{eqnarray}
CP|K_1^0\rangle \; &=& + |K_1^0\rangle \,, \nonumber\\
CP|K_2^0\rangle \; &=& - |K_2^0\rangle \, ,
\end{eqnarray}
a quantum number conserved in strong interactions.

Due to weak interactions, which are $CP$-violating, the kaons decay
and the ``physical'' states are the short and long lived states
\begin{eqnarray}\label{kaonSL}
|K_S\rangle \; &=& \frac{1}{N}\big\lbrace p |K^0\rangle - q
|\bar K^0\rangle \big\rbrace \,, \nonumber\\
|K_L\rangle \; &=& \frac{1}{N}\big\lbrace p |K^0\rangle + q
|\bar K^0\rangle \big\rbrace \, .
\end{eqnarray}
They are eigenstates of the non-Hermitian ``effective mass'' Hamiltonian.
In a particular phase convention, the weights are
expressed by \cite{Branco}
\begin{equation}\label{pqweights}
p=1+\varepsilon \, , \quad q=1-\varepsilon \, ,
\quad \mbox{and} \quad N^2=|p|^2+|q|^2 \, ,
\end{equation}
where $\varepsilon$ is the complex $CP$-violating parameter,
associated with the neutral kaon decay into the isospin 0 two-pion
state ($CPT$ invariance 
is assumed; thus the short and long lived states contain the same
$CP$-violating parameter $\varepsilon_S=\varepsilon_L=\varepsilon$).

Note that the two states $| K^0 \rangle$ and $| \bar K^0 \rangle$ can be
regarded as the quasi spin states up $|\Uparrow\rangle$ and down
$|\Downarrow\rangle$ and the operators acting in this quasi spin
space are expressible by Pauli matrices; the strangeness operator
$S$ can be identified with the Pauli matrix $\sigma_3$, the $CP$
operator with ($-\sigma_1$) and $CP$ violation in the effective
Hamiltonian is proportional to $\sigma_2$ \cite{BH1}.

Calculating now the probabilities of Eq.(\ref{UchiyamaBI}) within quantum mechanics
the Bell inequality (\ref{UchiyamaBI}) turns into an inequality for
the $CP$-violating parameter $\varepsilon$:
\begin{equation}\label{Uchiyamaepsilon}
\mbox{Re}\, \{\varepsilon\}\; \leq\; |\, \varepsilon\,|^2 \, .
\end{equation}
Inequality (\ref{Uchiyamaepsilon}) is obviously violated by the experimental value of
$\varepsilon \,$, having an absolute value of order $10^{-3}$ and a phase of about
$45^\circ$ \cite{ParticleData}. In this way $CP$ violation in $K^0 \bar K^0$ mixing is
related to the violation of a Bell inequality. 

Alternatively, we could choose $K_L$ instead of $K_S$ and $K_2^0$
instead of $K_1^0$ in the BI (\ref{UchiyamaBI}) and arrive at the same
inequality (\ref{Uchiyamaepsilon}). 

However, as already mentioned above, the derivation of inequality
(\ref{Uchiyamaepsilon}) relies on a specific choice of the phases of
the kaon states. In particular, the choice 
of the weights in Eq.(\ref{pqweights}), where the $CP$ violation parameter
$\varepsilon \,$ enters, is a convention such that the relative phase
of the decay amplitudes $K^0 \to \pi \pi$ and $\bar K^0 \to \pi \pi$,
both $\pi \pi$ states with isospin $I=0$, is $180^\circ$ (see, for instance,
Ref. \cite{Branco}). However, the BI (\ref{UchiyamaBI}) involves only
the two-dimensional space generated by the basis elements $| K^0
\rangle$ and $\bar K^0 \rangle$ and has nothing to do with
decays. This suggests to dispense with the phase convention
(\ref{pqweights}) and rather use the phase freedom to define the
unphysical state $| K^0_1 \rangle$.

This we can achieve by having a phase in the $CP$ transformation:
\begin{eqnarray}\label{CPtransformAlpha}
CP|K^0\rangle \; &=& - \, e^{i\alpha} |\bar K^0\rangle \,, \nonumber\\
CP|\bar K^0\rangle \; &=& -\,  e^{-i\alpha} |K^0\rangle \,,
\end{eqnarray}
where we have chosen $(CP)^2 = \mathbf{1}$.
In Eq.(\ref{CPtransform}) the phase $\alpha$ has been fixed for convenience to
$\alpha=0$, but in general it is arbitrary and without any physical significance. So the
$CP$ eigenstates are the following linear combinations
\begin{eqnarray}\label{CPstatesAlpha}
|K_1^0\rangle \; &=& \frac{1}{\sqrt{2}}\big\lbrace |K^0\rangle -
e^{i\alpha} |\bar K^0\rangle \big\rbrace \,, \nonumber\\
|K_2^0\rangle \; &=& \frac{1}{\sqrt{2}}\big\lbrace |K^0\rangle +
e^{i\alpha} |\bar K^0\rangle \big\rbrace \, ,
\end{eqnarray}
and with this definition the quantum mechanical
probabilities are
\begin{eqnarray}\label{QMprobab}
P_{QM}(K_1^0,\bar K^0) &=& \frac{1}{4} \,, \nonumber\\
P_{QM}(K_S,\bar K^0) &=& \frac{1}{2 N^2}|\, p|^2 \,, \nonumber\\
P_{QM}(K_S,K_1^0) &=& \frac{1}{4 N^2}|\, p\, e^{i\alpha} - q|^2 \, .
\end{eqnarray}
Note that besides $\alpha$ there is also the relative phase of $p$ and
$q$, which is still not fixed. 

We insert the probabilities (\ref{QMprobab}) into the Bell inequality
(\ref{UchiyamaBI}) 
and obtain 
\begin{eqnarray}\label{inequalphase}
\mbox{Re}\, \{e^{i\alpha} p\, q^*\} &\leq& |\, q|^2 \; .
\end{eqnarray}
Now we choose $\alpha$ such that it compensates the relative phase
$\chi$ of the complex weights $p$ and $q$:
\begin{equation}\label{alphafix}
\mbox{Re}\, \{e^{i\alpha} p\, q^*\} = \mbox{Re}\, \{e^{i(\alpha + \chi)}\,
|\,p|\, |\,q|\} = |\,p|\, |\,q| \, .
\end{equation}
Clearly, the inequality (\ref{inequalphase}) is optimal for $\alpha + \chi = 0$ and we
finally find an inequality independent of any phase conventions,
\begin{eqnarray}\label{inequalpq}
|\,p\,| &\leq& |\,q\,| \,.
\end{eqnarray}
Inequality (\ref{inequalpq}) is experimentally testable! Let us consider the semileptonic
decays of the $K$ mesons, in particular the leptonic asymmetry
\begin{eqnarray}\label{asymlept}
\delta_l &=& \frac{\Gamma(K_L\rightarrow \pi^- l^+ \nu_l) - \Gamma(K_L\rightarrow
\pi^+ l^- \bar \nu_l)}{\Gamma(K_L\rightarrow \pi^- l^+ \nu_l) + \Gamma(K_L\rightarrow
\pi^+ l^- \bar \nu_l)} \, ,
\end{eqnarray}
where $l$ represents either an electron or a muon. 
If $CP$ were conserved, we would have $\delta_l = 0$. Experimentally, however,
the asymmetry is nonvanishing\footnote{It is the weighted average over
electron and muon events, see Ref. \cite{ParticleData}.}, namely
\begin{equation}\label{deltaexp}
\delta_l = (3.27 \pm 0.12)\cdot 10^{-3} \, ,
\end{equation}
and is thus a clear sign of $CP$ violation.
On the other hand, we recall the $\Delta S = \Delta Q$ rule for the
decays of the strange particles. 
It implies that -- due to their quark contents -- the kaon $K^0(\bar s d)$
and the anti-kaon $\bar K^0(s \bar d)$ have definite decays
\begin{eqnarray}
K^0\stackrel{\bar s\rightarrow \bar u\, l^+ \nu_l} \longrightarrow \;
\pi^- + l^+ + \nu_l \; , \qquad
\bar K^0\stackrel{s\rightarrow u\, l^- \bar \nu_l} \longrightarrow \;
\pi^+ + l^- +\bar \nu_l \, .
\end{eqnarray}
Thus, $l^+$ and $l^-$ tag $K^0$ and $\bar K^0$, respectively, in the
$K_L$ state, 
and the leptonic asymmetry (\ref{asymlept}) is expressed by
the probabilities of finding a $K^0$ and a $\bar K^0$ in the $K_L$ state:
\begin{eqnarray}
\delta_l &=& \frac{|p|^2-|q|^2}{|p|^2+|q|^2} \equiv \delta \, .
\end{eqnarray}
Then inequality (\ref{inequalpq}) turns into the bound 
\begin{eqnarray}\label{inequaldelta}
\delta &\leq& 0 
\end{eqnarray}
for the leptonic asymmetry which measures $CP$ violation.
It is in contradiction to the experimental value (\ref{deltaexp}) which is
definitely positive. In this sense $CP$ violation is related to the violation
of a Bell inequality.

On the other hand, we can replace $\bar K^0$ by $K^0$ in the BI
(\ref{UchiyamaBI}) and along the 
same lines as discussed before we obtain the inequality
\begin{eqnarray}\label{inequalqp}
|\,q\,| &\leq& |\,p\,| \,,
\end{eqnarray}
independent of any phase conventions. The two inequalities (\ref{inequalpq}) and
(\ref{inequalqp}), however, imply the strict equality
\begin{eqnarray}\label{equalpq}
|\,p\,| &=& |\,q\,| \,,
\end{eqnarray}
which is in contradiction to experiment. Thus the premises of LRT
are only compatible 
with strict $CP$ conservation in $K^0 \bar K^0$ mixing. Conversely,
$CP$ violation in $K^0 \bar K^0$ mixing, 
no matter which sign the experimental asymmetry (\ref{asymlept})
actually has, always 
leads to a violation of a BI, either of inequality (\ref{inequalpq}) or of
(\ref{inequalqp}).

\vspace{6mm}

Another interesting feature is the connection of the Bell inequality with the decoherence
approach, see Ref. \cite{H1}. With a simple modification of the
quantum-mechanical probabilities, namely by 
multiplying the interference term of the amplitudes by $(1 - \zeta)$,
where $\zeta$ is the decoherence parameter, we can achieve a
continuous factorization of the wavefunction 
(see, e.g., Refs. \cite{BG,BGH1}). When does this approach represent a
local realistic theory, thus satisfying a Bell inequality? For $\zeta
= 0$ we have pure QM, the violation 
of a BI and thus a nonlocal situation. On the other hand, for $\zeta = 1$, called Furry's
hypothesis \cite{Furry}, there is total decoherence or spontaneous factorization of the
wavefunction. Then the BI is satisfied and a LRT may describe the physical phenomena.

However, what can we say for $\zeta$ values between $0$ and $1$? 
Let us consider again the
Bell inequality (\ref{UchiyamaBI}) and recalculate it with the modified probabilities,
in order to find a bound on $\zeta$. 
We choose for the entangled state the $K_S K_L$ basis (\ref{entangledKS})
and modify the probabilities as described above:
\begin{eqnarray}
& &P(f_1,f_2) = \frac{N^4}{8 |p|^2|q|^2}\,
\left| \langle f_1|_l\otimes\langle f_2|_r\; 
\biggl\lbrace|K_S\rangle_l\otimes|K_L\rangle_r - |K_L\rangle_l\otimes|K_S\rangle_r
\biggr\rbrace \right|^2 \nonumber\\
& &\longrightarrow P_\zeta(f_1,f_2) = \frac{N^4}{8 |p|^2|q|^2}
\biggl\lbrace|\langle f_1|K_S\rangle_l|^2 |\langle f_2|K_L\rangle_r|^2 +
|\langle f_1|K_L\rangle_l|^2 |\langle f_2|K_S\rangle_r|^2\nonumber\\
& &\qquad\qquad\quad
-\; 2\, (1-\zeta)\, \mbox{Re}\, \big\{\langle
f_1|K_S\rangle^*_l\langle f_2|K_L\rangle^*_r 
\langle f_1|K_L\rangle_l\langle f_2|K_S\rangle_r\big\}\biggr\rbrace \,.
\end{eqnarray}
Then we find the following probabilities modified by $\zeta$:
\begin{eqnarray}\label{zetaprobKS}
P_\zeta(K_1^0,\bar K^0) &=& P_{QM}(K_1^0,\bar K^0) \;-\; \zeta \,\frac{1}{8}\,
(1 - \eta^2) \,, \nonumber\\
P_\zeta(K_S,\bar K^0) &=& P_{QM}(K_S,\bar K^0) \;-\; \zeta \,\frac{1}{4}\,
(1 - \eta^2) \,, \nonumber\\
P_\zeta(K_S,K_1^0) &=& P_{QM}(K_S,K_1^0) \;+\; \zeta \,\frac{1}{8 \eta^2}\,
(1 - \eta^2)^2 \,,
\end{eqnarray}
where 
\begin{equation}
\eta = \frac{|q|}{|p|} 
\end{equation}
is a measure for $CP$ violation in $K^0 \bar K^0$ mixing.
Note that all $\zeta$ terms are independent of the phase $\alpha \,$; it enters only
in the quantum mechanical probability $P_{QM}(K_S,K_1^0)$ (see Eq.(\ref{QMprobab})).

Inserting now the probabilities (\ref{zetaprobKS}) into the Bell inequality
(\ref{UchiyamaBI}), choosing $\alpha$ -- like before in
Eq.(\ref{alphafix}) -- such that it compensates the relative phase
$\chi$ of the weights $p$ and $q$ and expressing 
$\eta$ by 
\begin{equation}
\eta^2 = \frac{1 - \delta}{1 + \delta} \,,
\end{equation}
we find the bound 
\begin{equation}
\frac{(1-\delta)}{\delta} (\sqrt{1-\delta^2} - 1 + \delta) \;\leq\; \zeta \,.
\end{equation}
The expansion to order $\delta$ gives
\begin{equation}
1 - \frac{3}{2} \delta \lesssim \zeta \,.
\end{equation}
Numerically, from the experimental value (\ref{deltaexp}) we get the bound
\begin{equation}
0.9951\pm 0.0002 \lesssim \zeta \, ,
\end{equation}
which is, due to our optimal choice of the phases in inequality
(\ref{inequalphase}), a
slight improvement as compared to the numerical bound of $0.987$ of
Ref. \cite{H1}. Thus, the
decoherence parameter $\zeta$ has to be very close to one; hence,
Furry's hypothesis or 
spontaneous factorization has to take place totally. Intuitively, we
would have expected that
there exist local realistic theories which allow at least partially
for an interference term, see for instance Refs. \cite{six2,selleri}.

On the other hand, we can compare this result with the experimentally
determined 
$\zeta^{K_S K_L} = 0.13^{+0.16}_{-0.15}$ (see Ref. \cite{BGH1}), where
$\zeta = 1$ is 
excluded by many standard deviations. This means that for experimental
reasons a LRT 
equivalent to a modification of QM in the $K_SK_L$ basis choice is
definitely excluded!

\vspace{6mm}

However, the situation changes when modifying the quantum-mechanical
probabilities in the $K^0 \bar K^0$ basis \cite{BG,BGH1}. 
Then we obtain
\begin{eqnarray}
P_\zeta(K_1^0,\bar K^0) &=& P_{QM}(K_1^0,\bar K^0) \,, \nonumber\\
P_\zeta(K_S,\bar K^0) &=& P_{QM}(K_S,\bar K^0) \,, \nonumber\\
P_\zeta(K_S,K_1^0) &=& P_{QM}(K_S,K_1^0) + \zeta \, \frac{1}{2 N^2} \,
\mbox{Re}\, \{e^{i\alpha} p\, q^*\} \,,
\end{eqnarray}
which implies with inequality (\ref{UchiyamaBI}) the lower bound
\begin{eqnarray}
1 - \sqrt{\frac{1-\delta}{1+\delta}} \;\leq\; \zeta \,.
\end{eqnarray}
To order $\delta$ we have
\begin{equation}
\delta \lesssim \zeta \quad \mbox{and numerically} \quad
0.0033 \pm 0.0001 \lesssim \zeta \; .
\end{equation}
Comparing this bound with the experimentally determined 
$\zeta^{K^0\bar K^0} \sim 0.4 \pm 0.7$ \cite{BGH1}, we see that we
cannot discriminate between QM and LRT in this case.

\vspace{6mm}

Summarizing, we have related Uchiyama's Bell inequality
(\ref{UchiyamaBI}) -- valid for the entangled $K^0 \bar K^0$ state
with negative $C$ parity -- with $CP$ violation in $K^0 \bar K^0$
mixing. Avoiding to involve any phase convention referring to $K^0$
and $\bar K^0$ decays, we have shown that Uchiyama's inequality
necessarily requires the $CP$-violating leptonic asymmetry $\delta$ to
be zero, in contradiction to experiment. In this way, $\delta \neq 0$
is a manifestation of the entanglement of the considered
state.\footnote{We want to stress that in the case of Uchiyama's 
Bell inequality (\ref{UchiyamaBI}), since it is considered at $t=0$,
it is rather  
contextuality \cite{noncontext} than nonlocality which is tested.} 
Amazingly, the non-zero result of $\delta$, obtained from measurements
at one-particle states, gives us information about the entanglement of
the two-particle state produced at the $\Phi$ resonance.
Moreover, connecting the
BI with the decoherence parameter $\zeta$, then the premises of
locality and reality 
are only compatible with a practically totally factorized
wavefunction, i.e., with $\zeta = 1$, and not 
with a partially contributing interference term.

\vspace{2cm}

\noindent {\large{\bf Acknowledgement}}\\This research was supported by the FWF Project
No. P14143-PHY of the Austrian Science Foundation and by the Project No. 2001-11 of the
Austrian-Czech Republic Scientific Collaboration.

\end{document}